# Production of Upgraded Metallurgical Grade (UMG) silicon for a low-cost high-efficiency and reliable PV technology

José Manuel Míguez Novoa[1], Volker Hoffmann[1], Eduardo Fornies[2], Laura Mendez[2], Marta Tojeiro[2], Fernando Ruiz[2], Manuel Funes[2], Carlos del Cañizo[3,*], David Fuertes Marrón[3], Nerea Dasilva Villanueva[3], Luis Jaime Caballero[3], Bülent Arıkan[4], Raşit Turan[4,5], Hasan Hüseyin Canar[4,6], Guillermo Sánchez Plaza[7].

[1] FerroGlobe, Spain

[2] Aurinka PV Group, Rivas-Vaciamadrid, Spain

[3] Instituto de Energía Solar, ETSI Telecomunicación, Universidad Politécnica de Madrid, Spain

[4] ODTÜ-GÜNAM (Center for Solar Energy Research and Applications), Turkey

[5] Middle East Technical University, Department of Physics, Turkey

[6] Micro and Nanotechnology, Middle East Technical University, Ankara, Turkey

[7] Nanophotonics Technology Center – Universidad Politécnica de Valencia, Valencia, Spain

**\* Correspondence:**
Corresponding Author: Carlos del Cañizo carlos.canizo@upm.es

**Keywords: photovoltaic solar energy, silicon feedstock, metallurgical route, solar grade silicon, solar cell manufacturing, module degradation, life cycle assessment**

**Abstract**

Upgraded Metallurgical Grade silicon (UMG-Si) has the potential to reduce the cost of Photovoltaic (PV) technology and to improve its environmental profile. In this contribution, we summarize the extensive work made in the research and development of UMG technology for PV, which has led to the demonstration of UMG-Si as a competitive alternative to polysilicon for the production of high-efficiency multicrystalline solar cells and modules. The tailoring of the processing steps along the complete Ferrosolar's UMG-Si manufacturing value chain has been addressed, commencing with the purification stage that results in a moderately compensated material due to the presence of phosphorous and boron. Gallium is added as a dopant at the crystallization stage to obtain a uniform resistivity profile ~1 $\Omega\cdot$cm along the ingot height. Defect engineering techniques based on phosphorus diffusion gettering have been optimized to improve the bulk electronic quality of UMG-Si wafers. Black silicon texturing, compatible with subsequent gettering and surface passivation, has been successfully implemented. Industrial-type Al-Back Surface Field (BSF) and Passivated Emitter and Rear Cell (PERC) solar cells have been fabricated, achieving cell efficiencies in the range of those obtained with conventional polysilicon substrates. TOPCon solar cell processing key steps have also been tested to further evaluate the potential of the material in advanced device architectures beyond PERC. Degradation mechanisms related to light exposure and operation temperature have been shown not to be significant in UMG PERC solar cells when a regeneration step is implemented, and PV modules with several years of outdoor operation have demonstrated similar performance to reference ones based on poly-Si. Life-Cycle Analysis (LCA) has been carried out to evaluate the environmental impact of UMG-based PV technology when compared to the poly-Si-based one, considering different scenarios both for the manufacturing sites and the PV installations.

# 1 Introduction

In 2000 the Spanish company Ferroatlántica, one of the key manufacturers of metallurgical silicon worldwide, started a research program for the production of solar grade silicon. Anticipating the silicon shortage already on the horizon due to the growth of the Photovoltaic (PV) market (Aulich and Schulze, 2002), Ferroatlántica (now Ferroglobe after the merge with the American company Globe Specialty Metals) aimed at upgrading its silicon to meet the requirements of the PV industry by applying "metallurgical methods", an interesting approach with the potential to reduce the energy consumption and associated costs, as compared to the incumbent Siemens technology. Ferroatlántica designed its own route based on the knowledge accumulated by the sector in the 70s-80s under several programs funded in USA, Japan and Europe after the oil crisis (Dietl, 1987). A small pioneer ingot was grown by the Institüt fur Kristallzüchtung (IKZ) with the first batch of upgraded silicon produced by Ferroatlántica. It was a compensated p-type material with resistivity of 0.015 $\Omega\cdot$cm due to the large amount of B contained in the feedstock, and it also incorporated a high amount of other impurities. Laboratory-type P/Al-BSF solar cells were fabricated at the Instituto de Energía Solar in Universidad Politécnica de Madrid to complete a whole manufacturing cycle, from metallurgical Si to solar cell fabrication. The recorded device efficiency was 10,7%, and the comparison of the Quantum Efficiency with that of a multicrystalline solar cell made from conventional electronic-grade polysilicon (Figure 1) clearly showed the limit imposed by the substrate quality (Del Cañizo *et al.*, 2005).

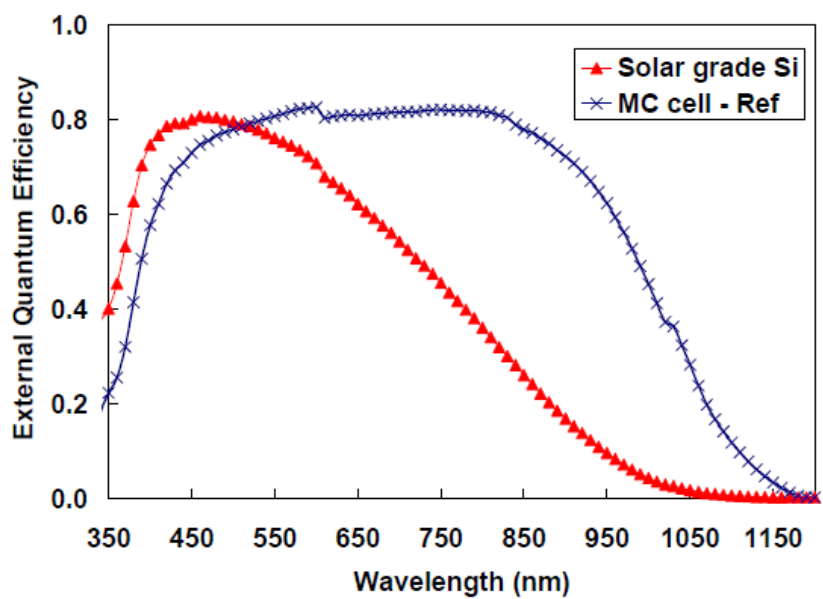

**Figure 1.** External Quantum Efficiency of a P/Al-BSF solar cell made with wafers from the first ingot grown with Ferroatlántica's solar grade silicon (red) in comparison with a reference made on conventional multicrystalline material (blue). Reproduced from (Del Cañizo *et al.*, 2005), with permission from EUPVSEC.

In 2017 Ferroglobe and Aurinka PV Group formed a joint venture called Ferrosolar to take advantage of acquired knowledge of metallurgical processes from Ferroglobe and the techno-economic experience from Aurinka in the solar industry. After a sustained R&D effort, industrial-type multicrystalline PERC solar cells fabricated on the solar silicon purified by Ferrosolar have already surpassed the 20% efficiency target, and PV modules with these cells have been in operation for several years demonstrating similar energy generation performance to that of neighboring reference modules made from conventional polysilicon (Forniés *et al.*, 2021).

The upgraded metallurgical technology was then ready for commercialization by the end of the last decade, and Ferrosolar planned to start industrial production of its solar grade silicon in Puertollano, Spain. Unfortunately, the initiative was stopped due to unfavorable market conditions. Despite this fact, the technology has successfully demonstrated its potential to reduce capital expenditure (CAPEX), manufacturing costs and energy consumption for the fabrication of PV-devices, as compared to those based on conventional polysilicon, and further developments are foreseen to keep up with the continuous progress that crystalline silicon technology is experimenting.

This paper updates the achievements of the solar grade silicon technology developed by the Ferrosolar project in Spain, reviewing its most relevant characteristics and assessing its environmental performance.

## 2 The Upgraded Metallurgical Silicon value chain, from feedstock to PV modules

The silicon feedstock purified by a sequence of metallurgical steps, thereby avoiding the chemical transformation into intermediate species (such as trichlorosilane in the dominant technology) has been referred to as Upgraded Metallurgical Silicon (UMG or UMG-Si) (Ceccaroli, Øvrelid and Pizzini, 2017). A number of initiatives advanced in the development of UMG-Si in the 2000s (Yuge *et al.*, 2001; Øvrelid *et al.*, 2006; Hoffmann *et al.*, 2008; Modanese *et al.*, 2011; Cocco *et al.*, 2013). Some of them reached the market but could not survive long, due to the reaction of the polysilicon industry that quickly expanded capacity and introduced processing optimizations, thus lowering manufacturing costs as compared to those achieved when serving just the microelectronic industry. Ferrosolar in Spain was among the main players following the UMG route. In the next sections the stage of technological development of its UMG-Si is reported, along with an analysis of its performance over the entire value chain, from the feedstock to the module.

### 2.1 Feedstock production

The main steps involved in the silicon purification process resulting in UMG-Si are sketched in Figure 2 and are described in detail in previous works (Hoffmann *et al.*, 2015; Forniés *et al.*, 2018). The process begins with the selection and cleaning of raw materials that will be reduced in the arc furnace to produce metallurgical silicon. The process concatenates then several steps that include slagging to reduce the high initial boron content, vacuum evaporation to reduce the similarly high initial phosphorus content, and the directional solidification to crystallise UMG-Si ingots and to simultaneously remove metallic impurities.

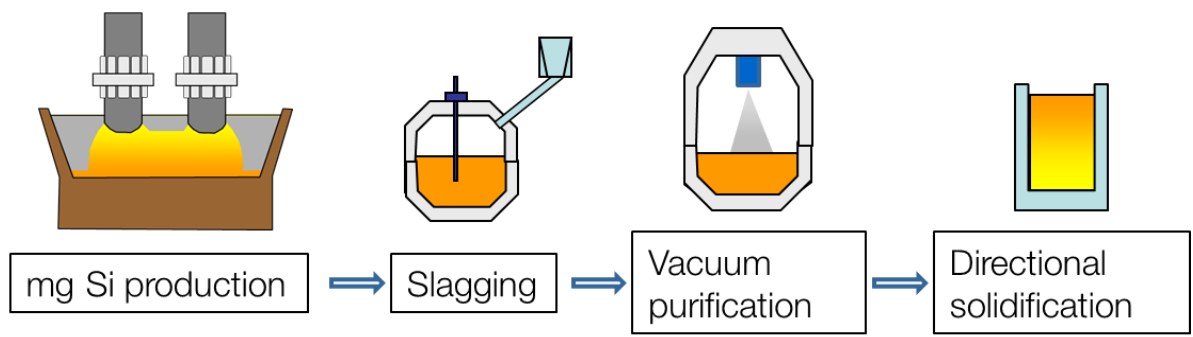

**Figure 2.** Sketch of the main steps in the upgrade of the metallurgical silicon. Reproduced from (Del Cañizo *et al.*, 2023) with permission from EUPVSEC.

The resulting UMG-Si product is characterized by a boron concentration below 0.2 ppmw, a phosphorus concentration in the range of 0.1–0.3 ppmw, and a total metal impurity concentration below 0.5 ppmw, which has been confirmed as a reasonable starting point for the production of competitive solar silicon feedstock.

## 2.2 Crystallization and wafering. Characterization before and after defect engineering

High Performance Multicrystalline ingots can be grown using this silicon feedstock. As both boron and phosphorus are present in UMG-Si, the different rate of incorporation from the liquid to the solid, due to their different segregation coefficients, would result in large variations of net doping concentration and electric resistivity along the ingot, and even a change of doping character, from p-type to n-type material, at the top of the ingot. This undesired feature should thus be corrected by adding controlled quantities of the additional dopants Ga (from 5 to 10 ppmw) and P (0.12 ppmw) to the melt. The impact of the dopant addition is exemplified in Figure 3(a), where the expected resistivity profile of two ingots, one without the addition of gallium and the other with it, are modelled. Figure 3(b) shows experimentally observed resistivity profiles of two different UMG-Si bricks grown at Pillar in Ukraine following this dopant-addition strategy.

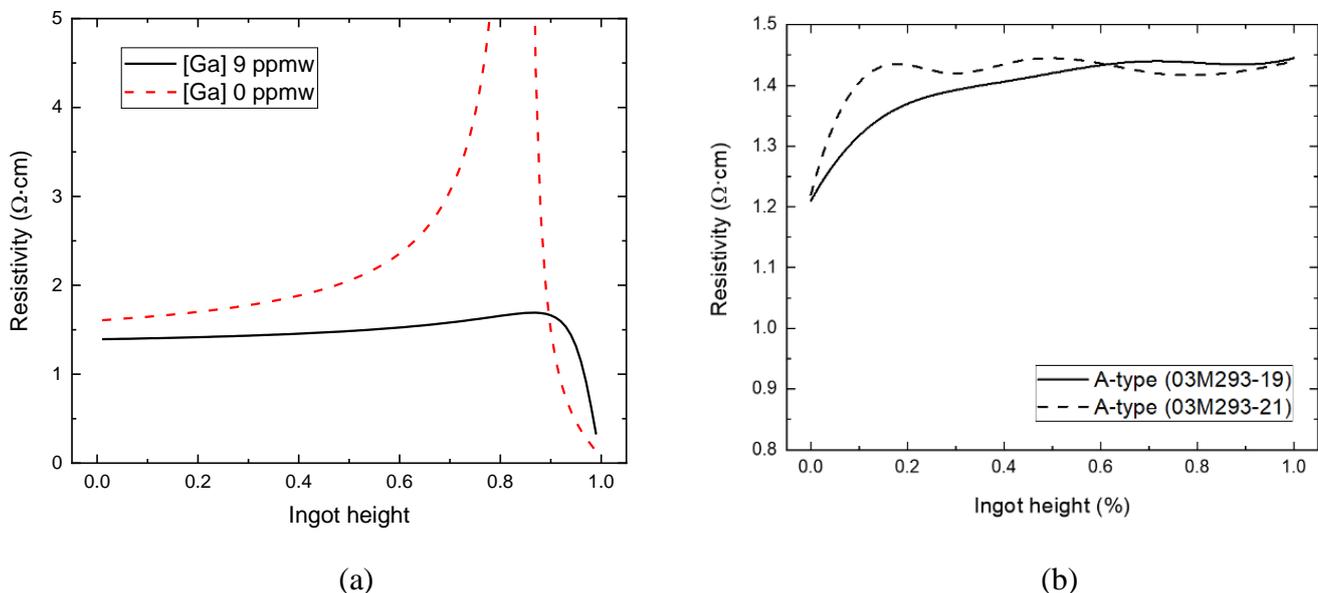

(a) (b)

**Figure 3.** (a) Calculated resistivity profiles applying Klaassen's model to two ingots with B and P concentration of 0.17 and 0.42 ppmw respectively, with no addition of Ga (dashed line) and adding 9 ppmw of Ga (straight line). (b) Experimental resistivity profiles measured in two UMG-Si bricks grown at Pillar with those concentrations.

Carrier lifetime maps of complete ingots measured with the Semilab WT-2000P tool are shown in Figure 4a. Discarding the top and bottom parts of each ingot due to high concentration of impurities, the measured average effective lifetimes are around 6 µs.

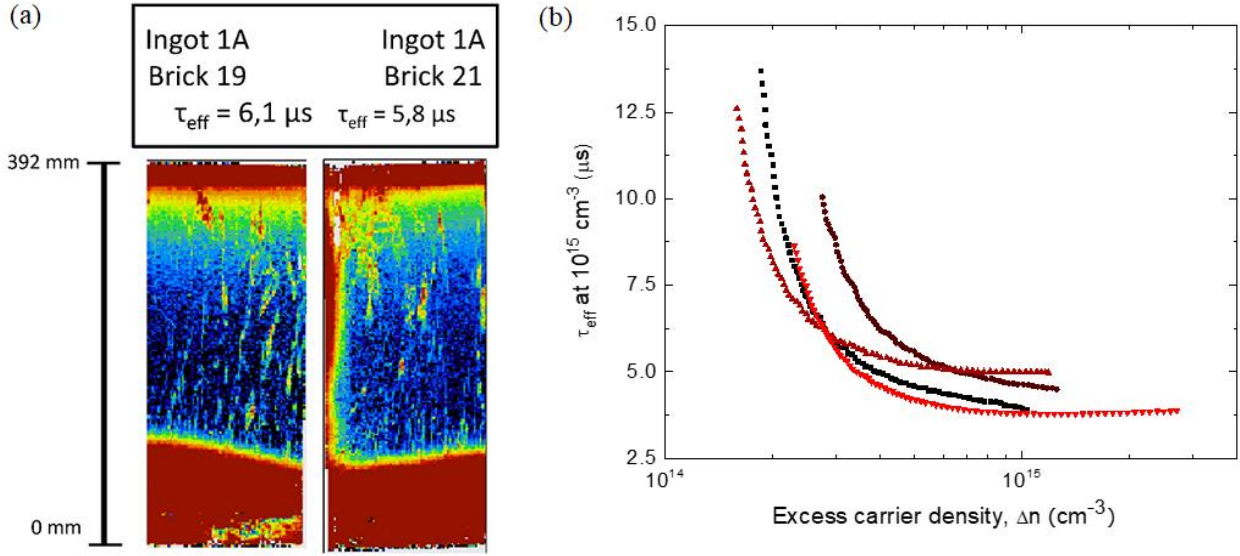

**Figure 4.** (a) Lifetime mappings of two UMG-Si ingots grown at Pillar as HPMC. Reproduced from (Forniés *et al.*, 2021) with permission from Elsevier. (b) Injection-dependent effective carrier lifetime of IE-passivated UMG-Si wafers.

Ingots should then be further cut into bricks and sliced into wafers. All the experimental tests below have been done on 156.75 × 156.75 mm$^2$ and 180 µm thick wafers.

Injection-dependent carrier lifetimes at wafer level have been measured by photoconductance decay (PCD) using the Sinton Instruments WCT-120 tool, after passivating the surfaces by immersion in a 0.1M iodine-ethanol (IE) solution. Very low lifetimes, in the 1 µs range, are typically obtained from bare wafers, as shown in **Figure 4**b, a clear indication that the material quality is limited by the presence of a large concentration of contaminant impurities. At low-injection levels, trapping effects commonly observed in multicrystalline materials are clearly seen. The impact of carrier trapping in UMG-Si has been analyzed in detailed elsewhere (Catalán-Gómez *et al.*, 2021; Dasilva-Villanueva *et al.*, 2022).

To effectively remove impurities from the bulk of the UMG-Si wafers, particularly those of metallic character, a phosphorous diffusion gettering (PDG) step is introduced prior to cell processing. For this purpose, a test was done selecting wafers randomly from different bricks cut from three different ingots, as to account for positional effects and batch variability. They were cut into 52 x 52 mm$^2$ samples, and firstly characterized with the Sinton tool, and then chemically treated with CP4 (HNO3/HF) etching, RCA1 ((HNH43OH/H2O2/H2O)) surface cleaning and a treatment in 2% HF to ensure native oxide removal before the thermal process.

P-diffusion was then performed in a tubular furnace, under different processing times (30, 60 and 90 minutes) and temperatures (780ºC, 800ºC, 820ºC and 850ºC). The diffusion was followed by a 10-minute drive-in at the same temperatures. The P-source was liquid POCl$_3$ kept at 26ºC, assuring super-saturation conditions during the diffusion step. After PDG, the emitter formed atop the surface was removed by chemical etching and PCD measurements were conducted again.

The impact of sequential PDG processes carried out under different conditions (duration and temperatures) on the bulk carrier lifetime of UMG-Si at a carrier injection level $\Delta n = 10^{15}$ cm$^{-3}$ can be seen in Figure 5. Best results for the first gettering process are systematically obtained at a "low" temperature (780ºC), resulting in lifetimes upgraded to an average level of 300 µs. Further improvements can be observed after a second PDG process, reaching an outstanding maximum lifetime of 722 µs. It is also interesting to note that this second P-diffusion provides an efficient gettering effect under a relatively wide range of temperatures, in particular and most conveniently, those compatible with P-diffusion processes for emitter provision and subsequent cell processing.

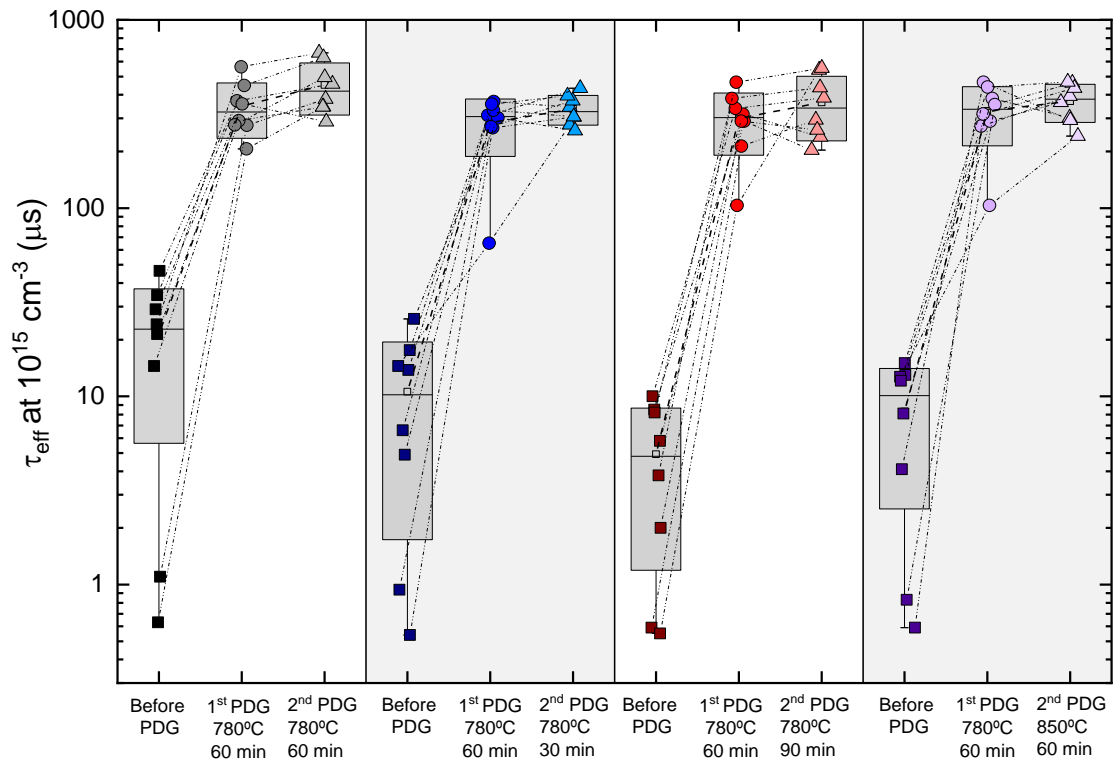

**Figure 5.** Effective lifetime variation before PDG, after one PDG process and after two PDG processes for different conditions of iodine-ethanol-passivated UMG-Si samples. In the box-type plot whiskers represent max-min variation, the height of the box represents the standard deviation, the non-filled square corresponds to the mean value and the middle line represents the median value. Connection dashed lines are included to identify the change of individual samples. Reproduced from (Del Cañizo *et al.*, 2023) with permission from EUPVSEC.

## 2.3 Solar cell development and results

The characterization program conducted on UMG-Si wafers led to conclude that they could feed a solar cell manufacturing process compatible with industrial upscaling. Three cell architectures have been tested for this purpose: Al-BSF, PERC and TOPCore. In the first two cases, large-area (156.75×156.75mm$^2$) solar cells were fabricated in industrially compatible production lines. In the third case, TOPCon key processing steps have been tested on the material to evaluate its potential beyond PERC.

### 2.3.1 Al-BSF solar cells

The first stage in the manufacturing of multicrystalline Al-BSF solar cells by a Tier-1 producer consists in a saw damage removal and a conventional acidic texturing. The P-emitter diffusion is performed in a low pressure diffusion furnace, and a back side chemical etching is performed to remove the diffused phosphorus from the back side, followed by an additional chemical etching for edge isolation. Thermal annealing follows to form a thin silicon dioxide to increase the resilience against Potential Induced Degradation (Luo *et al.*, 2017) of complete devices in operation, and subsequently the Silicon Nitride antireflective coating is deposited by Plasma Enhanced Chemical Vapor Deposition (PECVD). Front and rear contacts, as well as the Aluminum coating responsible for the Back Surface Field (BSF) formation, are printed in a double printing process, and a co-firing furnace is used to induce the drive-through of the contacts (Ballif *et al.*, 2002).

The performance of UMG-Si cells is compared in Table 1 with that of cells based on polysilicon wafers, processed in the same batch. Note that no tuning of the manufacturing parameters was implemented as to adapt the cell fabrication to the peculiarities of the incoming UMG-Si wafers; in particular, no additional PDG step was implemented, just the conventional diffusion to diffuse the P emitter.

**Table 1.** Average values of solar cell production. Comparison between polysilicon and UMG-Si for Al-BSF technology. The dashes in the cells stand for data not provided by the producer. Column Red. %rel refers to relative reduction in efficiency between UMG and poly. (Forniés *et al.*, 2019)

| Solar Cell Results | | $P_{MPP}$ [W] | $V_{oc}$ [V] | $I_{sc}$ [A] | FF (%) | Eta (%) | Red. (%$_{ref}$) | Counts |
|---|---|---|---|---|---|---|---|---|
| Al-BSF | Poly | 4.543 | 0.633 | 8.98 | 79.63 | 18.490 | 0.47% | - |
| | UMG | 4.522 | 0.632 | 9.02 | 79.27 | 18.404 | | 99692 |

### 2.3.2 PERC solar cells

An industrially-compatible process was designed for large-area PERC solar cells (**Figure 6**), optimizing the different steps for UMG-Si on pilot scale at ODTÜ-GÜNAM.

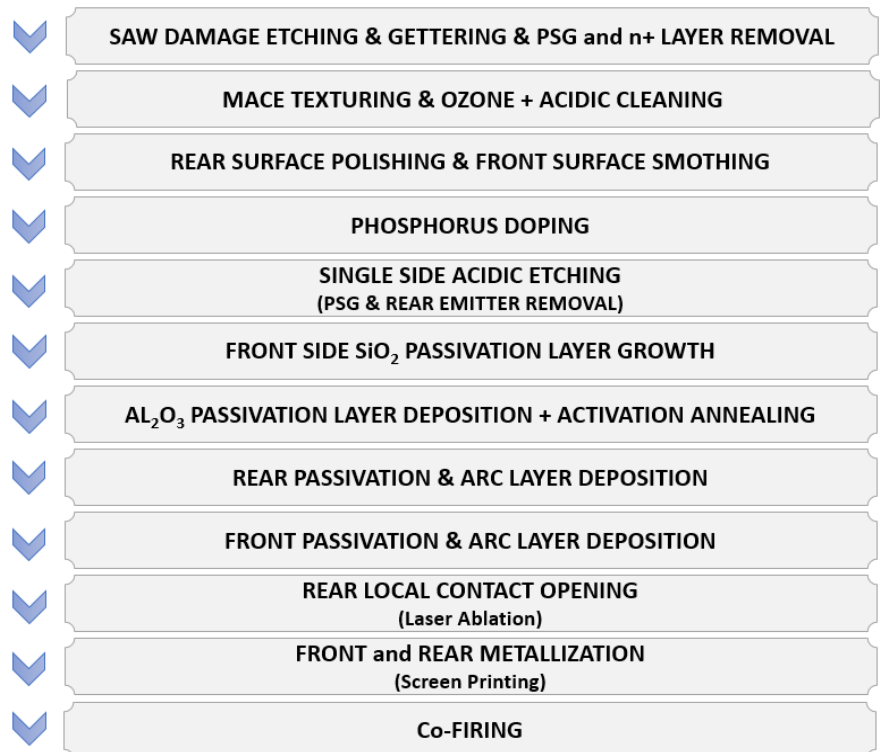

**Figure 6.** UMG-Si PERC solar cell process flow in an industrially compatible pilot line.

After saw damage removal in an alkaline bath, a PDG is implemented, as suggested by the optimization carried out in section 2.2, after which the wafers are symmetrically etched to remove the phosphosilicate glass (PSG) and the diffused emitter.

Metal (Silver)-Assisted Chemical Etching (MACE) is implemented to texture the wafers. The appearance of the UMG-Si wafer surface before and after MACE nanotexturing is shown in **Figure 7**. As it can be observed, the overall surface reflectance has decreased significantly after the treatment and the surface looks almost black. A SEM image of the MACE nanotextured surface is shown in **Figure 7**c.

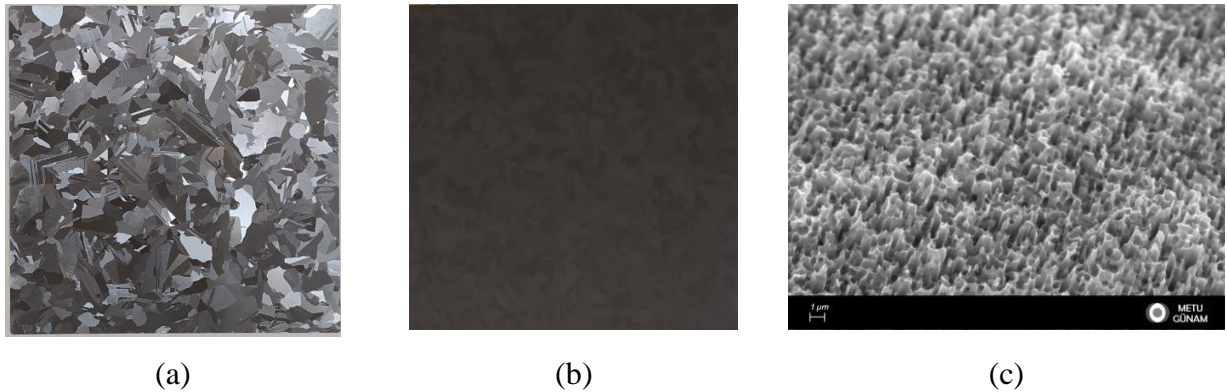

(a)    (b)    (c)

**Figure 7.** (a) Saw damage etched mc-UMG wafer, (b) mc-UMG wafer after MACE, (c) SEM image of MACE nanotextured surface.

The highly textured surface must then be prepared for the subsequent creation of a high-quality, effectively-passivated emitter by means of a chemical acidic etching step. This step represents a delicate tradeoff between a high aspect ratio and good surface passivation, and proceeds with a gentle smoothing of the nm-size surface features (thus leading to slightly increased surface reflectance) that permits an enhanced passivation efficacy. The treatment also eliminates the texturing from the cell rear side.

After the smoothing process, grains with different orientations show relatively different reflectance. A comparison between measurements done before and after smoothing on different grains is shown in **Figure 8**. The average reflectance value was determined in the range 11.5-12%, similar to reflectance values obtained from pyramidal alkaline texturing. As it can be observed, the optical response at short wavelengths following the MACE approach is improved compared to the standard pyramidal approach.

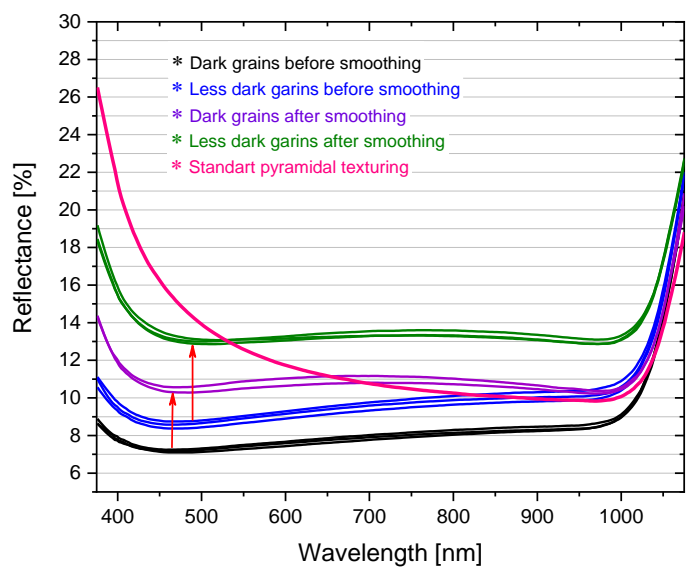

**Figure 8.** Reflectance measurements on different grains of MACE textured mc-UMG wafers before and after the chemical smoothing step. The reflectivity of a sample with standard pyramidal texturing is shown as a reference.

A phosphorus emitter was subsequently diffused from a POCl3 source in a tubular furnace, testing some recipes that resulted in sheet resistances in the range of 80-100 Ω/□. A single-side etch was then performed to remove the phosphosilicate glass (PSG) and the rear diffused region.

Different passivation layer stacks were tested both on MACE textured n+ doped front surfaces and polished p-type rear surfaces, to finally reach the structures sketched in **Figure 9**. As for the front surface, a stack formed by a hydrogenated amorphous silicon oxynitride (a-SiOxNy:H) layer and hydrogenated amorphous silicon nitride (a-SiNx:H) gave the best results, provided a thin silicon oxide layer was previously grown at low temperature (600ºC) on the emitter. A 2-stack version (**Figure 9**(a)) and a 4-stack one (**Figure 9**(b)) tuning indexes of refractions and layer thicknesses are implemented. As for the rear side, Atomic Layer Deposition (ALD) of thin alumina was chosen as the first passivating layer, after which a 20 nm hydrogenated amorphous silicon oxynitride (a-SiOxNy:H) layer plus a 100 nm hydrogenated amorphous silicon nitride (a-SiNx:H) capping layer were deposited by Plasma Enhanced Chemical Vapor Deposition (PECVD).

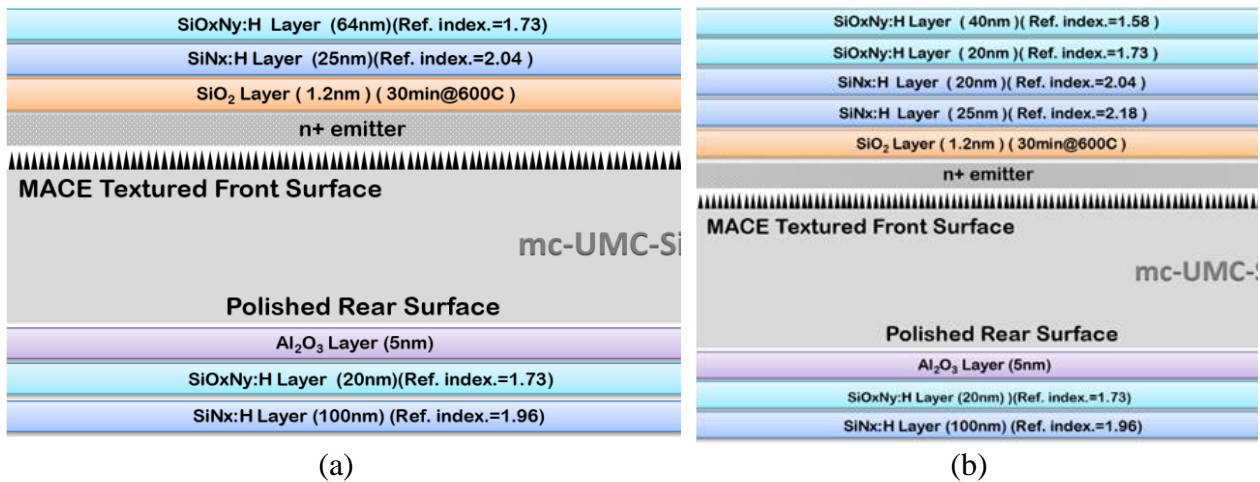

(a)          (b)

**Figure 9.** Proposed passivation structures for the front and rear surfaces of UMG PERC cells. (a) 2-stack on a SiO2 layer. (b) 4-stack on a SiO2 layer.

Rear local contact opening was done via laser ablation, covering a 3.19% fraction of the rear surface, which was determined as optimum from both simulation and experimental studies.

Regarding the screenprinting metallization and firing, a 5 busbar, 129 finger silver metallization pattern was applied on the front surface, and a full aluminium metallization at the rear. Problems arose in the form of discontinuities and narrowing of the metal fingers (see **Figure 10**) due to the use of pastes developed and optimized for pyramidal textured surfaces, and would be solved with new generation pastes and masks to be developed for nanotextured surfaces.

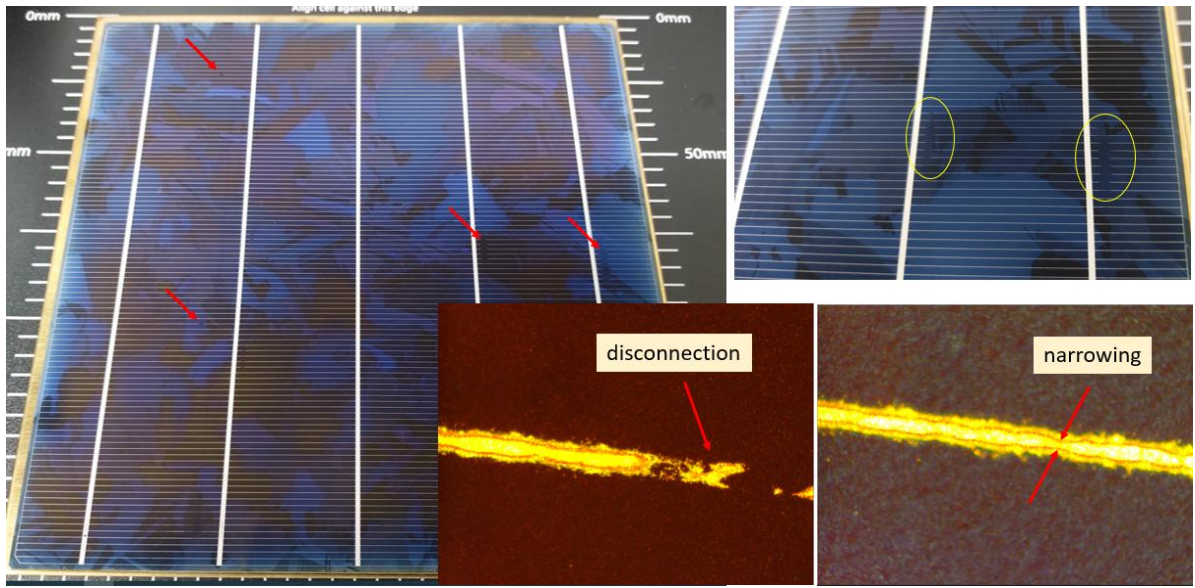

**Figure 10.** Major metal printing problems identified on the MACE-textured PERC cells.

Solar cell parameters measured after completion under standard test conditions are shown in **Figure 11**. Thanks to PDG and the rear and front passivation layer stacks, $V_{oc}$ values as high as 659 have been reached, in line with values routinely achieved with conventional polysilicon-based multicrystalline PERC cells in the same production line. Maximum $J_{sc}$ and FF values read 38.2 mA/cm$^2$ and 80.3%, respectively, slightly lower than those typically obtained with commercial industrial PERC cells and attributed to the use of pastes that are not optimized for nanotextured surfaces, as already explained, also responsible of the relatively wide distribution of values observed. Notwithstanding, a top cell efficiency of 20,1% has been achieved, a mark that could be further improved with the use of appropriate metal pastes and the implementation of additional advanced features, such as selective emitters or floating busbars, for example.

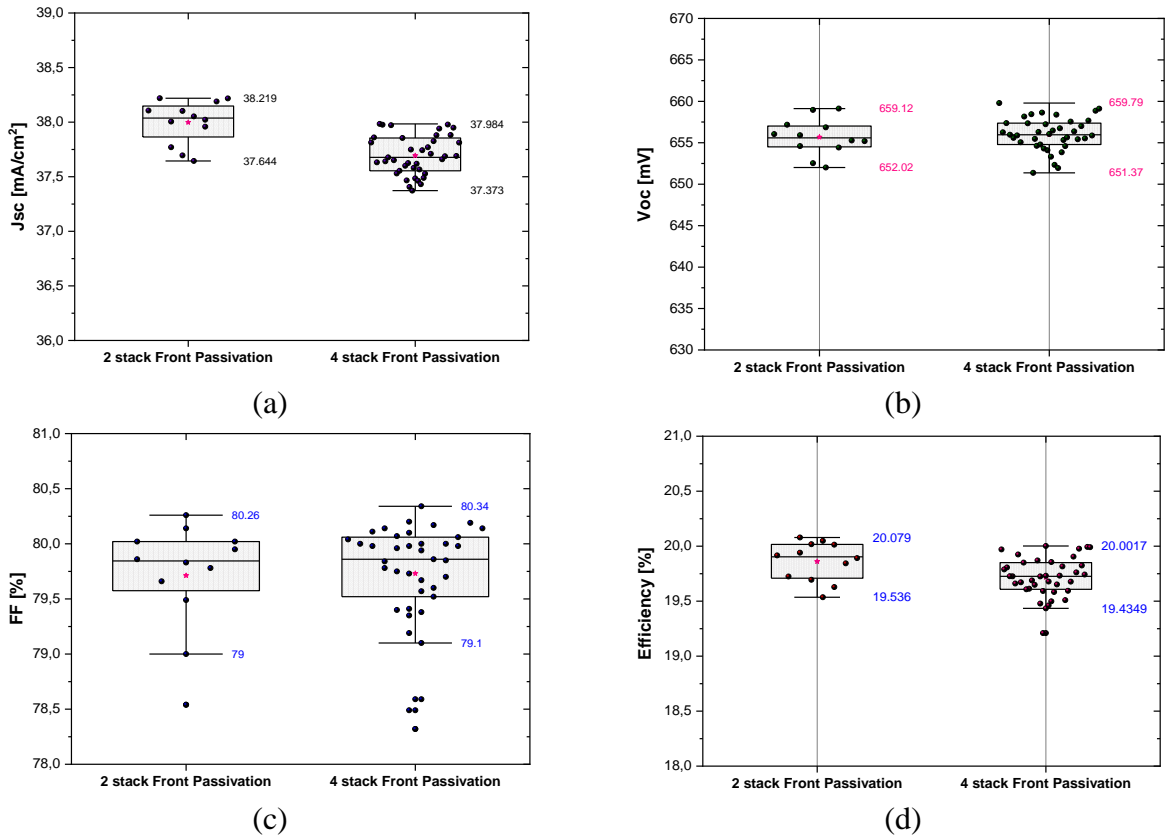

**Figure 11.** UMG-Si PERC solar cell parameters processed at ODTÜ-GÜNAM pilot line, measured under Standard Test Conditions. (a) Short circuit current density ($J_{sc}$). (b) Open Circuit Voltage ($V_{oc}$). (c) Fill Factor (FF). (d) Efficiency. Reproduced from (Del Cañizo *et al.*, 2023) with permission from EUPVSEC.

Additionally, the impact of light-induced degradation (LID) and light and elevated temperature induced degradation (LeTID) mechanisms has been evaluated on these cells. I-V and SunsVoc measurements of the fabricated cells were done after firing and the samples were kept under light (1 sun) at room temperature for controlled degradation. Regeneration was then performed on the degraded cells under optimized conditions (2 hours under 1 Sun light intensity at 140°C). Regarding LID, in controlled tests a degradation of up to 5 mV in $V_{oc}$ from the initial value was observed, but applying a light and heat assisted regeneration process, the Voc degradation was successfully reverted and the initial values were restored. As for the LeTID, for the time being no long-term tests have been completed yet, but short-term tests have shown no measurable effect.

Similar results have been obtained by a Tier-1 solar cell producer in its industrial pilot line, with the processing of a large number of UMG-Si multi wafers (more than 45.000) following no specific tuning of production parameters as to adapt them to UMG material. No pre-gettering step was included in this case. After saw damage removal, a black silicon nanotexture is performed by Reactive Ion Etching (RIE). Phosphorus diffusion, back side emitter removal, edge isolation and annealing followed, similarly to the previous case. The characteristic back surface passivation of PERC cells was implemented, consisting of a thin $Al_2O_3$ film plus a $SiN_x$ capping layer, both deposited by PECVD. Backside openings through the passivation stack were made by a nanosecond pulsed laser, and the screen-printing metallization followed. To minimize the light at elevated temperature degradation (LeTID) a post-treatment of the cells is performed.

Average parameters of the manufactured solar cells are shown in **Table 2**, comparing those obtained with UMG-Si wafers and those with polysilicon wafers processed in the same test. The record efficiency (not shown in the table) is 20.8 %.

**Table 2.** Average values of solar cell production. Comparison between polysilicon and UMG-Si for PERC technology. Column Red. %rel refers to relative reduction in efficiency between UMG and poly. (Forniés *et al.*, 2019)

| Solar Cell Results | | $P_{MPP}$ [W] | $V_{oc}$ [V] | $I_{sc}$ [A] | FF (%) | Eta (%) | Red. (%ref) | Counts |
|---|---|---|---|---|---|---|---|---|
| B-Si (RIE) + PERC | Poly | 5.014 | 0.7 | 9.62 | 80.02 | 20.410 | 1.37% | 55396 |
| | UMG | 4.945 | 0.6 | 9.54 | 79.87 | 20.130 | | 46197 |

Some of these PERC solar cells have been tested for LeTID following the IEC TS 63342 norm ('IEC TS 63342 — C-Si photovoltaic (PV) modules — Light and elevated temperature induced degradation (LETID) test — Detection', 2022), to assess whether the preventive treatment applied to avoid it is indeed effective. As depicted in **Figure 12**, $V_{oc}$ readings reflect an initial decrease right after current injection and a gradual recovery of the cell voltage afterwards. The cell was preconditioned to avoid the influence of B-O complexes during the test, and then measured to determine its output power. Once the treatment was stopped, the power was measured again, confirming a power reduction below 3% with respect to the starting values. This result leads to conclude that the preventive treatment was successful in making the mc-UMG-based PERC cell

LeTID-insensitive.

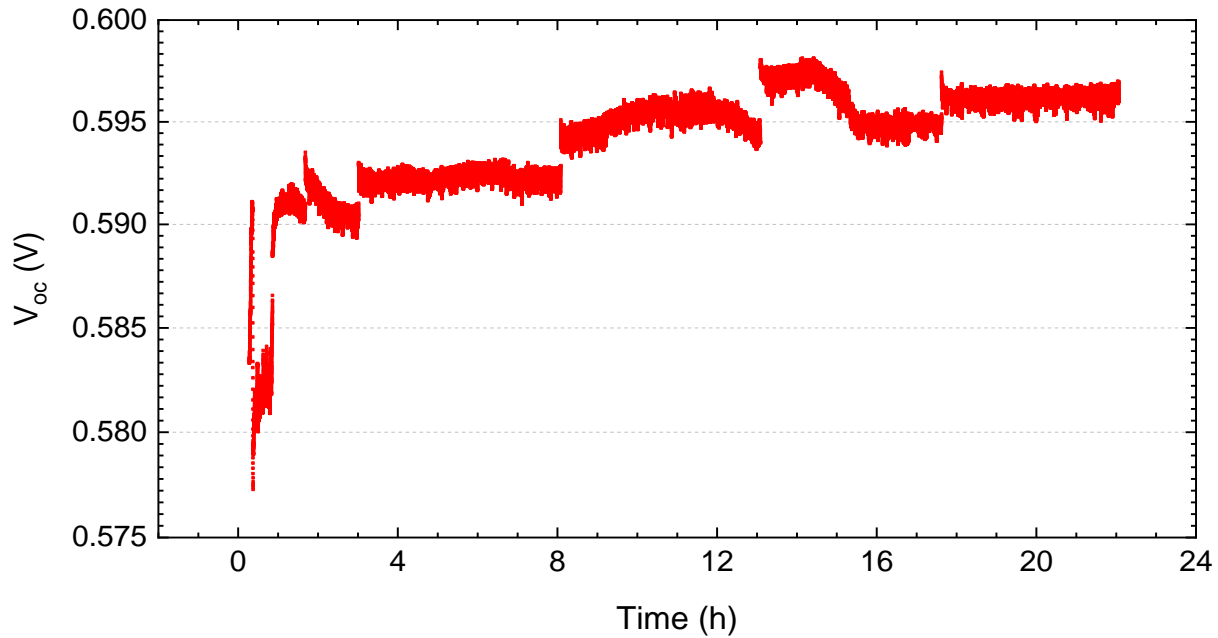

**Figure 12.** $V_{OC}$ evolution of a UMG-Si PERC solar cell under the treatment conditions of injected current and temperature.

### 2.3.3 TOPCoRE solar cells

To further assess the potential of UMG to keep up and follow the rapid evolution of silicon technology, additional experiments were performed at Fraunhofer ISE to implement TOPCon rear emitter structure, (TOPCoRE) (Richter *et al.*, 2021) on the UMG-Si substrates.

As a first step, bulk lifetimes after PDG and TOPCon-compatible annealing were measured by calibrated Photoluminescence (PL), confirming carrier lifetime values in the range of hundreds of microseconds from most of the area of the processed wafers, even locally surpassing the millisecond in some regions.

The samples were satisfactorily plasma-textured, achieving a reflectance (without ARC) below 5.9% in the range 300 to 1000 nm. An excellent surface passivation could be reached on these textured surfaces by means of $Al_2O_3/SiN_x$ layer stacks, yielding implied $V_{oc}$ >700 mV in good areas of the processed wafers.

Based on injection-dependent PL results, an efficiency limiting bulk recombination analysis (ELBA) has been conducted in order to infer local PV-efficiency distributions over the probed sample area (Michl *et al.*, 2012). As shown in **Figure 13**, a global efficiency above 22% is within reach with these multicrystalline UMG-Si substrates.

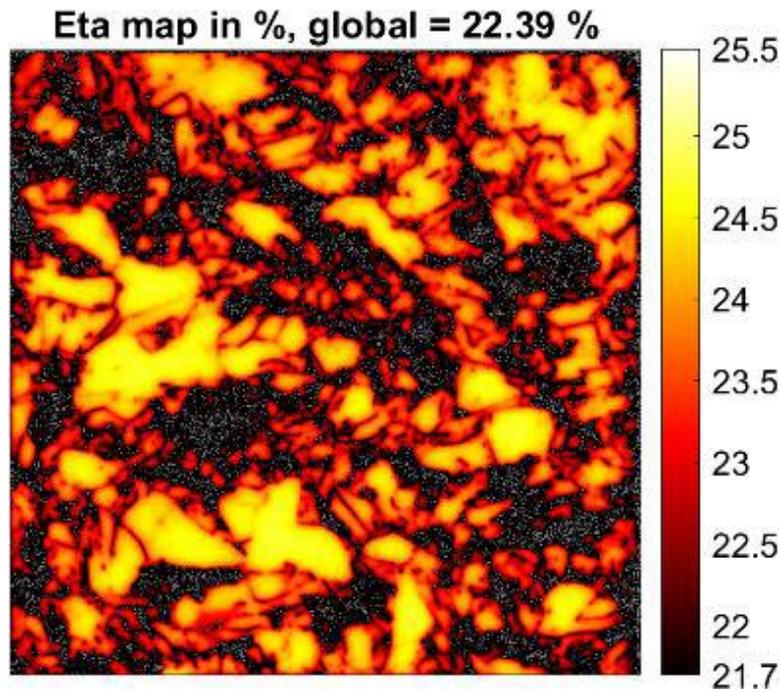

**Figure 13.** Estimated efficiency of UMG-Si solar cells according to ELBA analysis for a TOPCore structure. Reproduced from (Del Cañizo *et al.*, 2023) with permission from EUPVSEC.

### 2.4 Modules

Both Al-BSF and PERC UMG-Si-based solar cells processed in industrial lines have been further assembled into PV modules, following a standard production process including string formation and tabbing, interconnection, lamination and finishing. As an example, the average module performance of 72-cells modules are reported in **Table 3.** Average values of module production (Forniés *et al.*, 2019).

**Table 3.** Average values of module production (Forniés *et al.*, 2019).

| Cell Technology | $P_{MAX}$ [W] | $V_{oc}$ [V] | $I_{sc}$ [A] | $V_{MPP}$ [V] | $I_{MPP}$ [A] | FF (%) | Counts |
|---|---|---|---|---|---|---|---|
| Al-BSF | 325.1 | 45.73 | 9.32 | 37.13 | 8.76 | 76.31 | 1350 |
| Black Si + PERC | 354.4 | 47.78 | 9.70 | 38.21 | 9.28 | 77.00 | 300 |

### 3 In-field operation of UMG PV modules

PV modules with UMG-Si cells have already been installed in a variety of places around the world. Here we will report on some controlled tests done over large time periods. A demonstration photovoltaic system with UMG-Si modules was installed in Ávila, Spain (Sánchez *et al.*, 2011). The modules were mounted on a fixed structure with an inclination of 45º and south orientation. This PV-system consists of three generators of 3.3 kW peak each, two of them with multicrystalline UMG-Si modules of 60 and 72 Al-BSF cells and nominal power ratings of 230 W and 270 W, respectively, and the third one with commercial multicrystalline silicon modules of 60 cells and 230W which served as a reference. Each of the three generators is connected to a 3.3kW inverter provided with two maximum power point (MPP) trackers. In this way, the PV-system is configured with six strings, each one connected to an MPP-tracker.

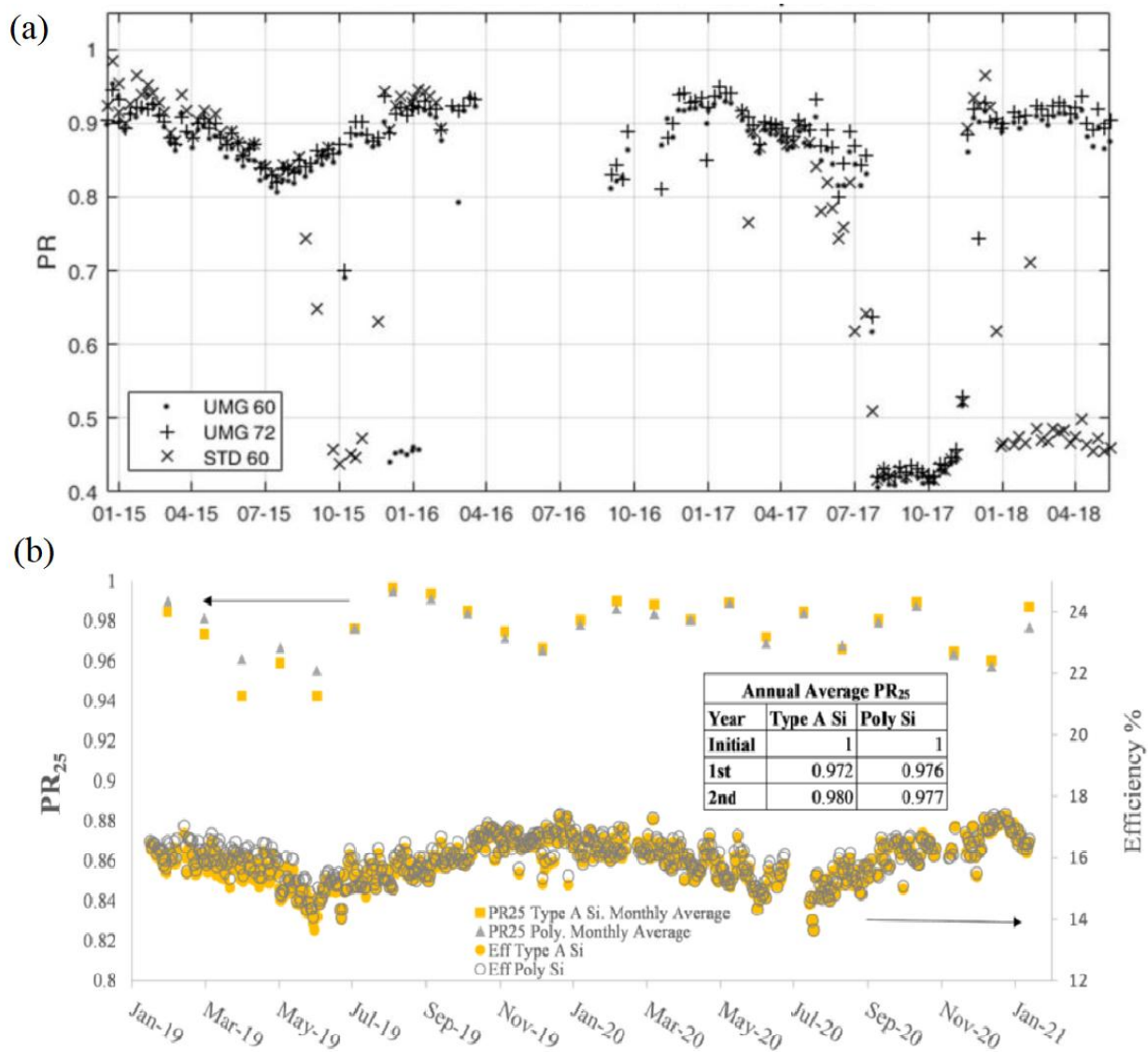

**Figure 14**. (a) Recorded PR of the three generators installed in Avila, Spain, two of them provided with mc-UMG-Si modules of 60 and 72 cells and the third one with commercial 60 mc-Si cells (STD 60). PR ranges between 0.4 and 0.5 correspond to periods operating with a single string due to issues related to the inverters. Reproduced from (Sánchez *et al.*, 2011) with permission. (b) Module efficiency and monthly average of PR calculated at 25 ◦C for UMG-Si (Type A) and polysilicon recorded over 24 months in operation. Missing data correspond to interventions on the equipment. Reproduced from (Forniés *et al.*, 2021), with permission from Elsevier.

**Figure 14**a shows the evolution of the performance ratio (PR) of the system, excluding partial shading intervals, during an operational period of 3 years, with each point representing one week time lapse. PR values ranged between 0.8 and 0.9 most of the time. The evolution observed is very similar for the three different generators, with a marked seasonal behavior. The PR is lower

in summer due to the higher operating temperature of the modules and the lower rainfall. As it can be observed, three years after commissioning of the plant, the UMG modules did not show any noticeable degradation over the reference modules based on standard poly-Si.

Another demonstration plant was installed more recently in Tudela (Spain), with 7 UMG-Si modules with Al-BSF cells installed in the same rack together with 7 commercial modules made of polysilicon-based solar cells (Forniés *et al.*, 2021). Both generators were mounted on a fixed structure with a 30° tilt with respect to the horizontal plane, south-oriented, and completely shadow-free. Each generator was connected to the grid via a 2.5 kW inverter, always operating at MPP. The output power, module temperature and incoming irradiance were continuously monitored over 24 months of operation, so as to calculate the module efficiency and the monthly averages of the performance ratio at 25°C (PR25), which are shown in **Figure 14**b. Results clearly show a very similar behavior of both types of modules (UMG-Si and poly-Si). The inset in **Figure 14b** shows the annual PR25 average values over two years of test. After the first year, the PR25 decreased from 1 to 0.972 for UMG-Si and to 0.976 for poly Si, which is attributed to the initial light induced degradation (LID) typical of Al-BSF solar cells (Pingel *et al.*, 2010). After 2 years, the recorded PR25 was even higher for UMG-Si modules (Type A, PR25 = 0.980) than for polysilicon modules (PR25 = 0.977).

These results have been further corroborated by Guerra et al (Guerra *et al.*, 2022), who calculated the degradation rates of both types of modules over the two-year period of the study by means of two different procedures, leading to exactly the same result: a mean degradation of around 0.2%/year in the reference poly-Si-based modules (SoG label in the figure), while no measurable degradation was found for UMG-Si modules.

The remarkable conclusion of this work was that no evidence of the existence of any additional degradation mechanism that may eventually affect UMG-Si material could be found.

## 4 Environmental analysis

To evaluate the differences in the environmental impact associated to the manufacturing of UMG-Si-based cells with respect to other cell technologies based on standard polysilicon, a process-based Life-Cycle Analysis (LCA) has been performed, according to the Methodology Guidelines on LCA of Photovoltaic Electricity published by the International Energy Agency (IEA) and following the Environmental Footprint (EF) 3.0 impact assessment method, as proposed by the European Union (EU) (Official Journal of the European Union, 2013; Fazio *et al.*, 2019; R. Frischknecht *et al.*, 2020).

Note that this kind of analysis of environmental impact goes beyond the purely scientific and technological relevance, reaching out other aspects of societal, political or economic importance. Nevertheless, it turns out that the environmental impact associated to UMG technology is a key aspect justifying its interest, and it is therefore relevant to compare UMG Si vs. polysilicon in this regard.

In order to proceed with the analysis, it is necessary to refer production to specific energy mixes at relevant production sites, as the electric power input required for manufacturing has a decisive role in the overall impact of the PV value chain (Yue, You and Darling, 2014). For this purpose, national electricity mixes have been typically used in the literature and they will be similarly used here, even though considering them as variables of the study may be controversial and subject to dynamic evolution in short terms. In the following, the reference to specific electricity mixes should not be simply read as representative of the geographical region itself, but rather of different constitutional shares of the respective mixes (fossil fuels, renewables, nuclear, etc.). Four electricity mixes are used as referent mixes in this section, including dissimilar proportions of primary generation technologies. The Spanish (ES) electricity mix is use as the example of the major proportion of renewable electricity, while the presence of coal is residual. The European Union (EU), followed by the United States (US) and China (CN), are used as examples of mixes with increased ratios of non-renewable sources or fossil fuels. European countries are, in average, characterized by around 20% of nuclear energy, which is slightly reduced in the case of US, while for China's mix it is almost residual. Likewise, the share of coal generation for US and EU is around 15-20% while in China it counts for more than 60% of the total generation.

The software Simapro 9.0 has been used, with Ecoinvent 3.5 as database for all the background data (Wernet *et al.*, 2016). Additionally, crystalline silicon solar cell processes have been updated, recurring to up-to-date literature when appropriate.

## 4.1     Environmental analysis of PV systems with UMG-Si Al-BSF modules

For the Al-BSF technology, the LCA covered the whole PV value chain, starting from metallurgical silicon production down to PV-electricity generation, using as a functional unit 1 kWh of electricity produced by a ground-mounted multicrystalline PV system with a nominal capacity of 100 MWp installed in the South of Spain, whose key parameters are summarized in **Table 4**. The complete study can be consulted in (Méndez *et al.*, 2021).

**Table 4.** Key parameters for the comparison of the LCA of a PV system with multi BSF modules, either with UMG or polysilicon substrates.

| Module efficiency | Module degradation rate | PV lifetime | BoS lifetime | Annual irradiation | Inclination | Performance ratio | Time frame of data |
|---|---|---|---|---|---|---|---|
| UMG: 18.43% | 0.40% | 30 years | 15 years electrical gear | 2160 kWh/m2 year | Optimal tilt angle 34° | 82.50% | 2015-2020 |
| Poly: 18.55% | | | 30 years structure | | | | |

Among the most important results obtained from the study, see **Figure 15a**, is the verification that the usage of UMG-Si instead of conventional polysilicon can lead to an overall 20% reduction in climate-change-related emissions when considering the Spanish (ES) mix (12.10 vs. 14.95 gCO$_2$eq/kWhe), which is even higher for the Chinese (CN) mix (16.12 vs. 21.38 gCO$_2$eq/kWhe).

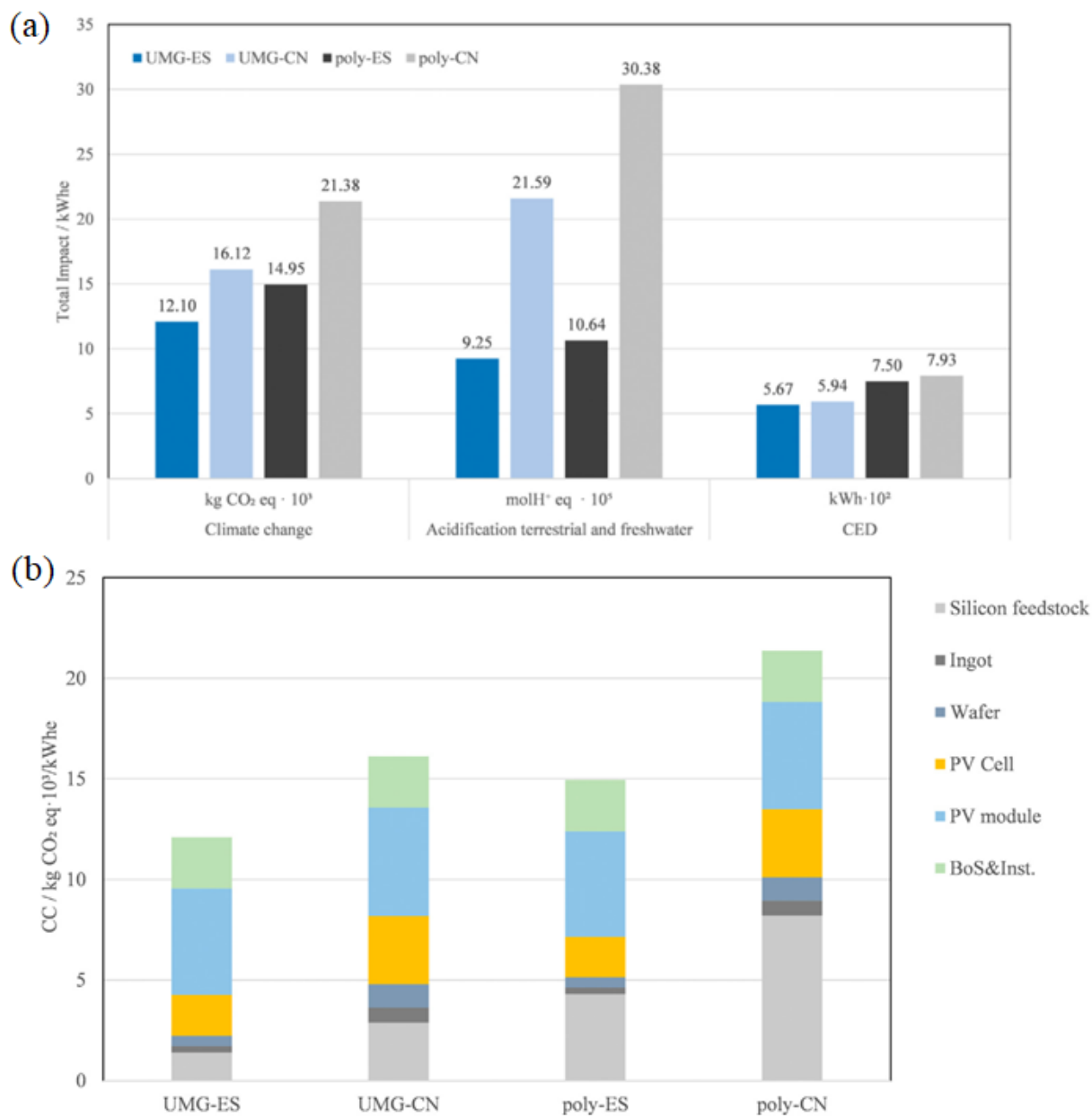

**Figure 15.** (a) Main impacts per kWh of PV-produced electricity for the different scenarios: Spain (ES) vs China (CN) electricity mix, and UMG-Si vs poly-Si material manufacturing. All cases consider Al-BSF cell technology. Reproduced from (Méndez *et al.*, 2021) with permission from Elsevier. (b) Climate Change contributions of different stages of the production of PV systems per kWh of produced electricity, for UMG and polysilicon BSF technologies, when the manufacture is located in Spain (ES) and China (CN). Reproduced from (Méndez *et al.*, 2021) with permission from Elsevier.

**Figure 15a** complements these results with the impact on Acidification: Terrestrial and Freshwater (ATF) and the calculation of the total cumulative primary energy demand (CED). The Energy Pay Back Time (EPBT) of the system, which can be deduced from the latter (Louwen *et al.,* 2015), shows similar trends to other impacts, being halved for the case of UMG-Si-based modules manufactured with the Spanish mix as compared to the case of poly-Si-based technology manufactured in China.

**Figure 15b** shows the corresponding shares of the different manufacturing stages of a full PV-system to the Climate Change emissions for the different scenarios considered. As it can be observed, the most significant differences affect the silicon feedstock and the cell fabrication stages, with an important impact of the electricity mix considered. In all cases, UMG-Si outperforms poly-Si for a given mix.

### 4.2 $CO_2$-payback time of UMG-Si Al-BSF technology.

The composition of the electricity mix at the location of the actual UMG-Si-based PV-system is relevant to calculate the $CO_2$-payback time, *i.e.*, the time required for the system to offset the greenhouse gases emitted as a result of its construction so that it begins saving GHG emissions as compared to the conventional grid-based electricity. The following calculations want to compare this payback time for the UMG-Si and polySi based systems characterized in the previous section, considering exclusively parameters affecting the PV-generator (those module technology- and manufacturing-dependent) and not including others, such as the balance of system.

The 100 MW photovoltaic generator described in section 4.1 is used as a reference, locating it in different European places: Almeria (South-East Spain), Marseille (South-East France), Landes (South-West France), Paris (France), Lille (France) and Frankfurt (Germany). These locations have been selected as representative of three different national electricity mixes, with different levels of related emissions: low for France (58 g$CO_2$eq/kWh), medium for Spain (165 g$CO_2$eq/kWh), and high for Germany (348 g$CO_2$eq/kWh), according to data published by the European Environment Agency ([www.eea.europa.eu](www.eea.europa.eu)) for 2021. As additional input data, the annual average irradiation (kWh/m$^2$/year) at the given locations has been considered. The $CO_2$-payback time results are summarized in **Table 5**.

**Table 5.** Estimation of the $CO_2$-payback time of a reference PV system manufactured in China (CN) and Spain (ES) and located in different European sites.

**Location analysis**

| Location | Irradiation (kWh/m²/year) | gCO₂eq/kWh | | CO₂ payback time (years) | |
|---|---|---|---|---|---|
| | | Poly-Si CN | UMG-Si ES | Poly-Si CN | UMG-Si ES |
| Almería (SE-Spain) | 2150 | 21.4 | 12.1 | 2.4 | 1.4 |
| Marseille (SE-France) | 1650 | 27.9 | 15.8 | 14.3 | 8.1 |
| Landes (SW-France) | 1350 | 34 | 19.3 | 17.5 | 9.9 |
| Paris (France) | 1150 | 40 | 22.6 | 20.5 | 11.6 |
| Lille (France) | 1050 | 43.8 | 24.8 | 22.5 | 12.7 |
| Frankfurt (Germany) | 1000 | 46 | 26 | 3.1 | 1.8 |

The last two columns in the table show the calculated payback time (in years) necessary to counteract the emissions incurred during fabrication of the PV-modules. As it can be observed, UMG-Si, fabricated in a medium-rate-emissions mix like that of Spain, systematically yields lower payback times (~56%) in comparison to poly-Si modules fabricated in China. The time span is obviously largely influenced by the solar resource at the site, but significantly more (about 10 times, comparing the cases in France and Germany) by the actual mix to which the PV-system is injecting energy.

## 4.3 Comparison of the environmental impact of different cell technologies

As for the environmental analysis of PERC and TOPCon structures, additional studies have been performed, focused only on cell technology and not on full PV-systems, in contrast to the previous study. With current cell-to-module factors very close to unity, the conclusions can be safely extrapolated up to UMG-Si PV module level. **Table 6** shows the different schemes of PV cells evaluated, among which both polysilicon and UMG-Si materials are considered. The table includes input data from different sources.

**Table 6.** Parameter values used for the evaluated scenarios of the environmental analysis, including different cell architectures.

| Work ref and cell technology | Silicon type | Wafer area (m²) | Cell eff (%) | Pmax (Wp/cell) | Pmax (Wp/m²) | Module eff (%) | $P_{MAX}$ (W) | Electricity consumption (kWh/m²) |
|---|---|---|---|---|---|---|---|---|
| (Forniés *et al.*, 2019), Al-BSF | poly-Si | 0.02457 | 18.50 | 4.55 | 185 | 16.72 | 324.0 | 18.31 |
| (Forniés *et al.*, 2019), Al-BSF | UMG | 0.02489 | 18.40 | 4.58 | 184 | 16.84 | 326.4 | 18.27 |
| (Luo *et al.*, 2018), PERC | poly-Si | 0.02457 | 19.20 | 4.72 | 191.9 | 17.34 | 336.2 | 24.42 |
| (R. Frischknecht *et al.*, 2020; Frischknecht, 2021), Al-BSF | poly-Si | 0.02457 | 18.80 | 4.62 | 188 | 16.99 | 329.3 | 17.7 |
| (Del Cañizo *et al.*, 2023), PERC | UMG | 0.02457 | 19.93 | 4.90 | 199.3 | 18.01 | 349.1 | 19.13 |
| (Del Cañizo *et al.*, 2023), TOPCoRe | UMG | 0.02457 | 22.40 | 5.50 | 224 | 20.24 | 392.3 | 20.35 |

As above, all categories considered in the EF methodology were calculated, with the focus now set on the following categories: Climate Change (CC), Acidification (ATF), Resource Use, Metals and Minerals, and Ionizing Radiation. Results are shown in **Figure 16**.

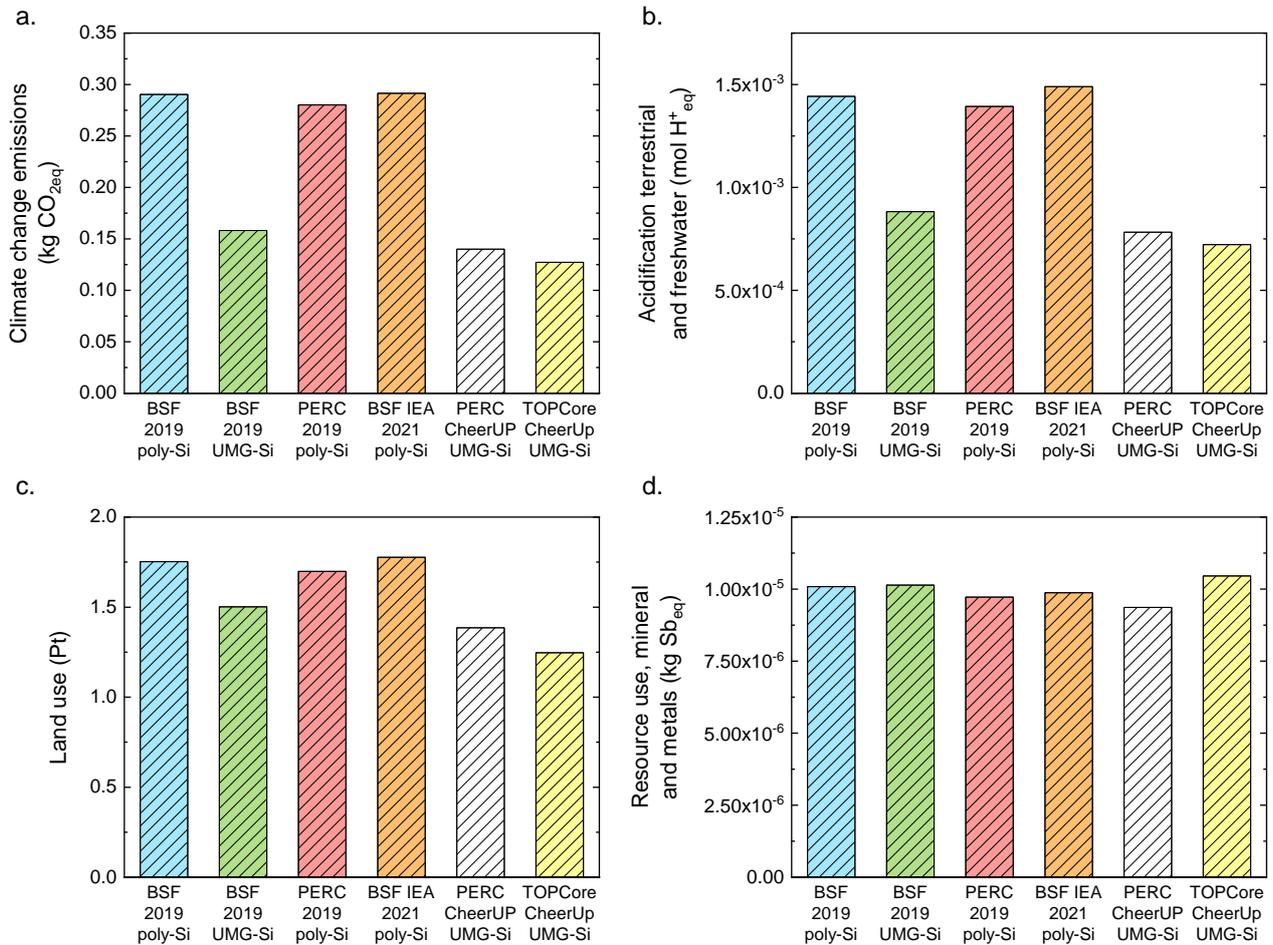

**Figure 16.** Cell-processing impacts (in standard units) for selected categories (ES mix scenario). (a) Climate change emissions in kgCO2eq per cell Wp. (b) Acidification terrestrial and fresh water in mol H+eq per cell Wp. (c) Land use in standard units. (d) Use of resources in kg Sb eq per cell Wp.

Climate change and acidification categories present the most significant differences when comparing silicon feedstocks. UMG-Si shows better performance on both parameters, mainly because its manufacturing process is less energy intensive, leading thus to lower associated emissions. In CC category UMG-Si is in the range of 0.158-0.127 kgCO$_2$eq per cell, while polysilicon cells generate between 0.292-0.280 kgCO$_2$eq. Within the CC category, the observed improvements are aligned with the progress in cell architectures and their improved efficiencies. Small differences are found between data sources for BSF cells, though. The acidification parameter shows a similar trend. Concerning the resource use, the results are quite similar, but the TOPCoRe architecture shows a slight increase due to greater necessities related to metallization. In general, TOPCon schemes incur in a larger use of silver and aluminium per peak watt than BSF and PERC.

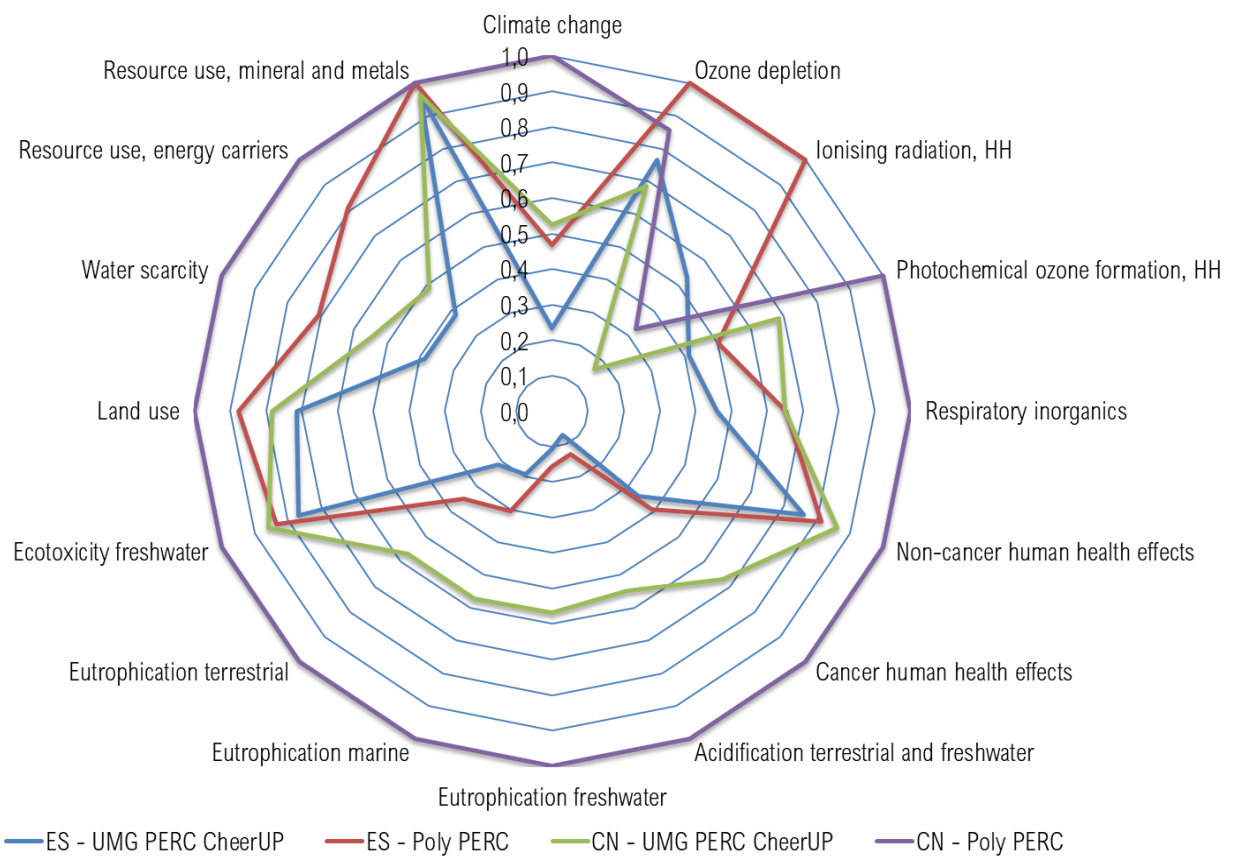

**Figure 17.** Normalized environmental impacts of polysilicon and UMG PERC technologies, comparing the manufacture in Spain (ES) and China (CN). Reproduced from (Del Cañizo *et al.*, 2023) with permission from EUPVSEC.

The trends for the normalized values in all EF categories for PERC architectures considering ES and CH mixes are summarized in **Figure 17**, revealing the strong influence of the manufacturing location over all the environmental categories considered. The predominant presence of fossil fuels, specifically coal, in the Chinese mix has a strong influence over these results. In the same way, it is concluded that for PERC cells the use of UMG-Si leads approximately to a 50% reduction in terms of CC emissions for both ES and CN mixes with respect to conventional poly-Si. However, for the ATF category, the reduction is 45% for the ES mix, and 76% for the CN mix. Also, it is interesting to notice how, for the CN mix, some categories suffer from sharper changes, as for example eutrophication or cancer human health categories. Therefore, the relevance of the silicon feedstock is even more meaningful when the energy mix contains reduced shares of renewables energies.

In order to show in more detail the environmental impact as a function of cell technology, substrate material and electricity mix for some of the most relevant categories, the results are summarized in **Figure 18**.

For the CC category (**Figure 18**(a)), cells made of poly-Si show values between 0.62-0.28 kg$CO_2$/Wp, while for UMG-Si cells they are 0.37-0.13 kg$CO_2$/Wp. Manufacturing cells in CN, US and EU mixes imply 54%, 48% and 16% increase of climate change emissions, respectively, with respect to the ES mix.

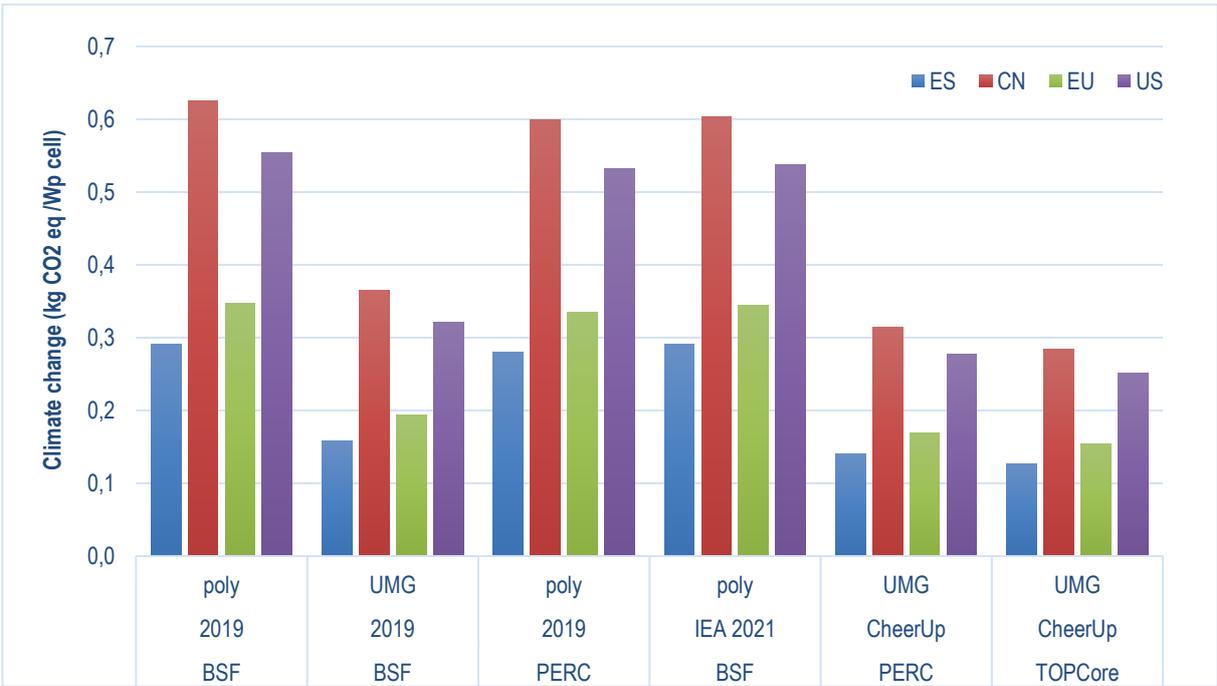

(a)

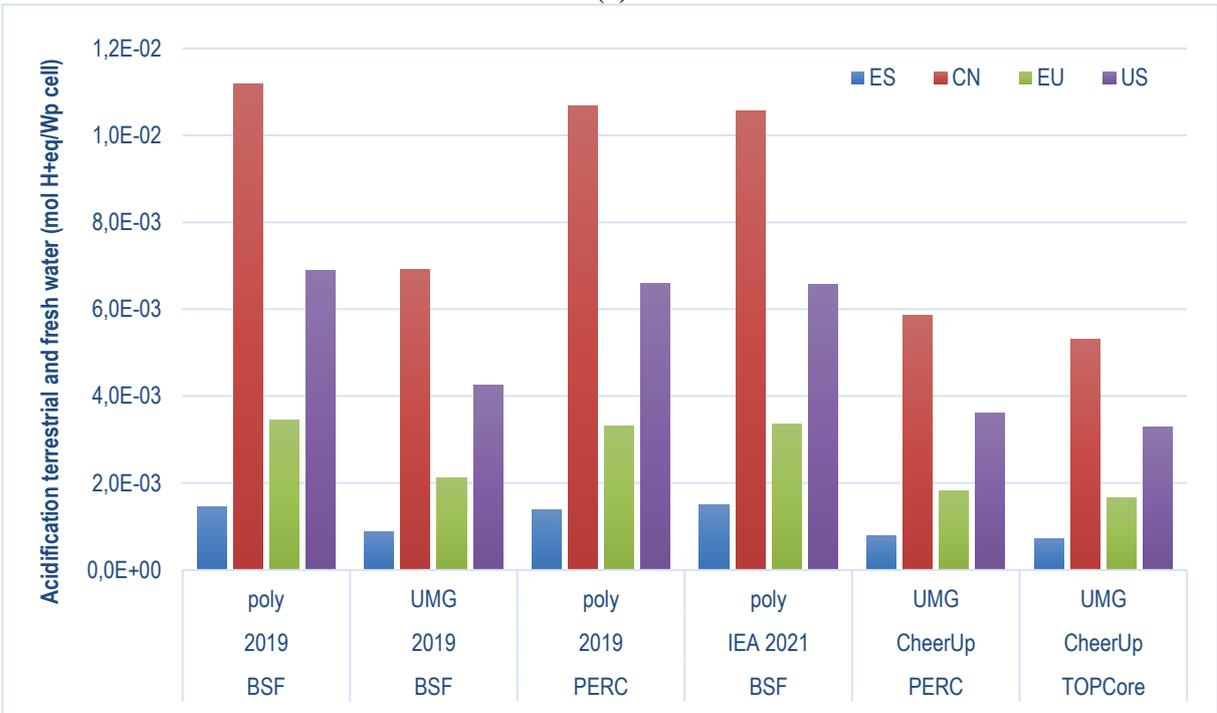

(b)

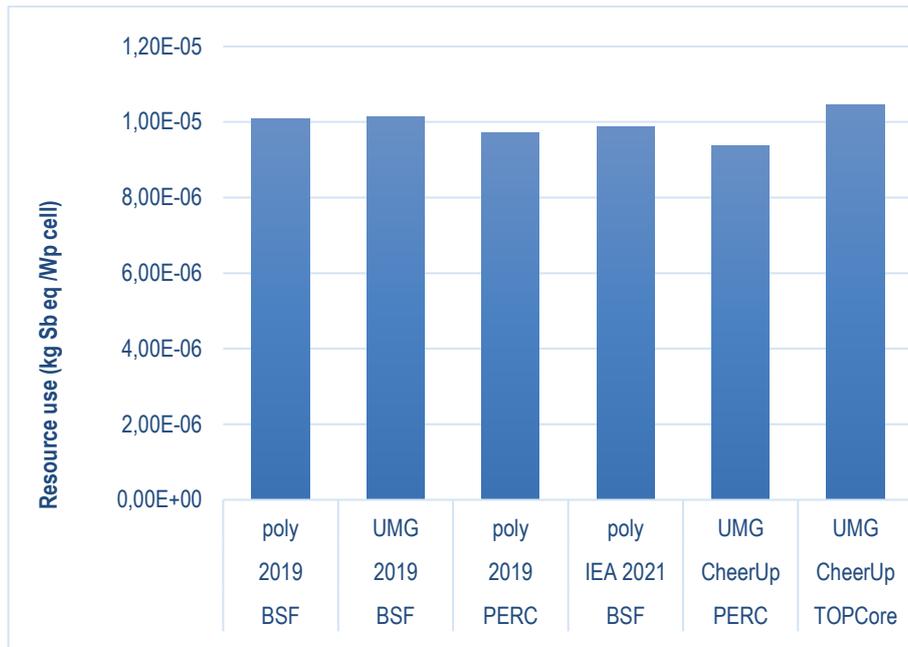

(c)

**Figure 18.** Comparison of environmental impact of conventional polysilicon and UMG-Si technologies for four electricity mixes: Spain (ES), China (CN), European Union (EU) and United States (US). (a) Climate change emissions in kgCO2eq per cell Wp. (b) Acidification terrestrial and fresh water in mol H+eq per cell Wp. (c) Use of resources in kg Sb eq per cell Wp.

Differences are even more important for the acidification (ATF) category (**Figure 18**(b)), on account of the almost negligible proportion of coal in the ES mix. Manufacturing cells in CN, US and EU mixes imply 87%, 79% and 58% higher acidification impact than if fabricated under the ES mix. It must be noticed that both PERC-UMG and TOPCoRe-UMG have an ATF-impact around 50% lower than Poly-BSF, for the same electricity mix. In the more uneven case, UMG-PERC cells show 79% and 93% less impact in CC and AFT categories beside Poly-BSF (IEA 2021), when ES and CN mixes are selected.

As for resource use of minerals (**Figure 18**(c)), the location has not such importance as for CC or AFT, since the differences among mixes are less than 0.01%. The use of minerals does not depend on energy mix or location, but only on cell technology. Therefore, among UMG cells, PERC and TOPCoRe technologies have a -7,7% and 3.1% of resource impact variation as compared to UMG BSF cells, mainly due to the necessities of metallization of each cell architecture, and the high levels of optimization achieved for the current PERC-cell processing. Contrasting different feedstocks, in this case UMG PERC and TOPCoRe technologies have a -5.2% and 5.9% of resource impact variation when compared to poly-BSF cells. On the other hand, the observed slight differences among BSF cells are probably just due to small parameter variations among different data sources.

As a summary, the grouped environmental impact contributions of UMG-Si PERC cell technology for the Spanish electricity mix scenario are shown in **Figure 19**.

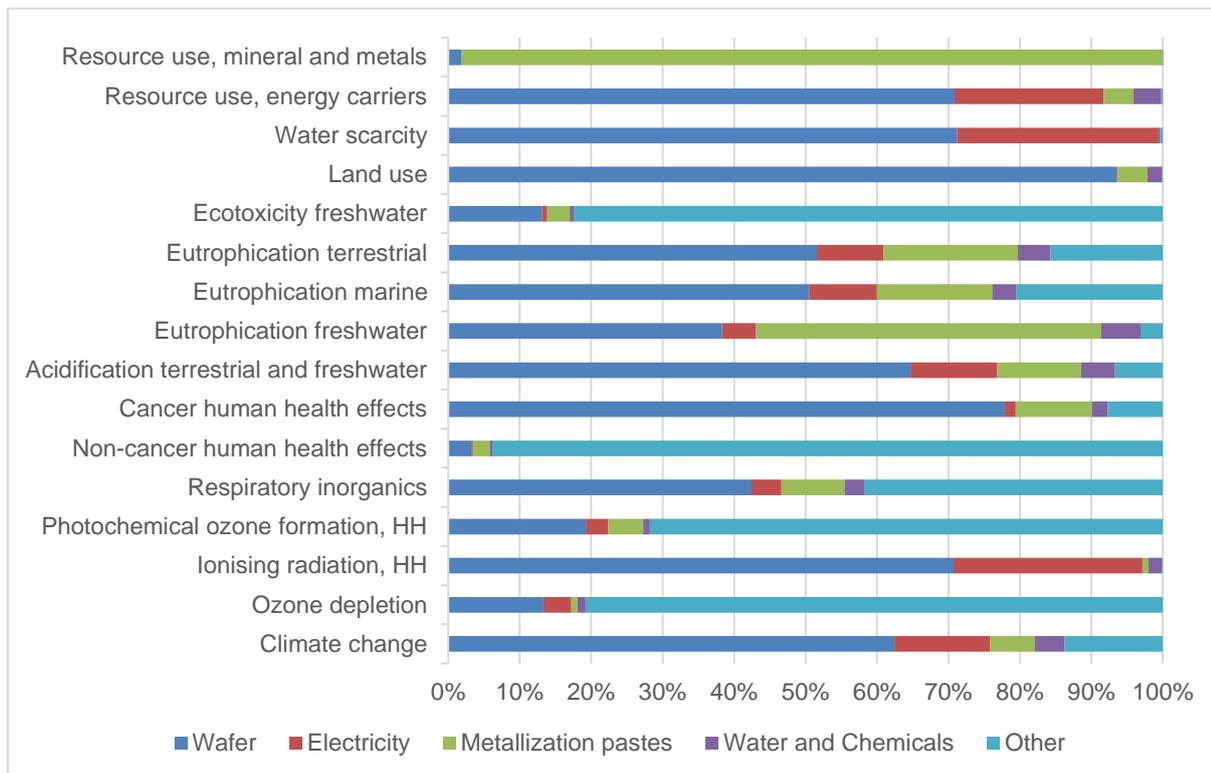

**Figure 19.** Grouped contribution to impact categories (ES – PERC – CheerUp – UMG)

As it can be observed, wafer manufacturing is the most relevant contributor to most environmental impacts, including those related to resource use (energy carriers), water scarcity, acidification, climate change or ionizing radiation. Electricity consumption in general has a lower impact than the wafer, but in some parameters, such as resource use (energy carriers), or ionizing radiation, it becomes more relevant. The electricity use during the cell production has not such a critical role as compared to wafer production. As it could be expected, the main contributions associated to the metallization pastes affect the resource use of minerals and metals, but also the eutrophication of water. The result is highly influenced by the consumption of metals, specifically silver, in the course of the metallization processes and the eventual contamination that could produce over freshwater.

## 5 Conclusions

A proprietary technology for the purification of silicon for solar applications has been developed and assessed along the entire value chain, from the feedstock up to the complete PV system.

The purification process involves a number of metallurgical steps (slagging, evaporation, solidification) to upgrade the purity of metallurgical silicon, resulting in the so-called Upgraded Metallurgical Silicon (UMG-Si). In this material impurities are present in concentrations compatible with further solar cell processing, the most prominent ones being boron, phosphorus and metals in concentrations below 0.2 ppmw, 0.3 ppmw and 0.5 ppmw, respectively. High Performance Multicrystalline ingots have been produced, showing uniform resistivities along the ingot height thanks to the addition of gallium, resulting in a compensated material of ~1 Ωcm. Initial bulk lifetimes, in the range of 10 µs, can boost to several hundred microseconds (with maximum of 722 µs) after an optimized Phosphorus Diffusion Gettering step.

The significant improvement in the UMG-Si electronic quality allowed the achievement of cell efficiencies in the range of those obtained with conventional polysilicon substrates for industrial-type Al-Back Surface Field (BSF) and Passivated Emitter and Rear Cell (PERC) multicrystalline technologies (18,4% and 20,1% as representative efficiency values, respectively). Furthermore, its potential for advanced solar cell technologies beyond PERC has been demonstrated, with bulk carrier lifetimes compatible with 22% device efficiency after TOPCore solar cell key thermal steps.

UMG Al-BSF PV modules have demonstrated similar performance to reference ones, maintaining PR25 over 0.97 after two years of outdoor operation. It has also been shown that a conventional regeneration step in manufacturing successfully inhibits light- and elevated-temperature induced degradation for UMG PERC solar cells.

The environmental impact of UMG technology has been shown to be lower than that of conventional polysilicon, irrespective of the manufacturing site or the cell technology employed, with significant reductions in climate-change emissions higher than 20% and simultaneous reductions in the energy pay-back time of 50%.

In summary, a sustained R&D effort deployed in the last two decades has succeeded in demonstrating that UMG-Si can be the basis of a low-CAPEX, low-cost environmental-friendly supply of PV silicon.

## 6    Acknowledgements

This work has been partially funded as part of the R&D SOLAR-ERA.NET Cofund 2 project CHEER-UP, funded by the Spanish MCIN/AEI/10.13039/501100011033/ through projects PCI2019-111834-2 and PCI2019-111903-2, by the Spanish Centro para el Desarrollo Tecnológico Industrial (CDTI), and by the Turkish TÜBITAK through Project 219M029. MCIN/AEI/ 10.13039/501100011033 is also acknowledged for financial support through the GREASE project (PID2020-113533RB-C31).